# SAFEGUARDING VOICE PRIVACY: HARNESSING NEAR-ULTRASONIC INTERFERENCE TO PROTECT AGAINST UNAUTHORIZED AUDIO RECORDING

Forrest McKee[1] and David Noever[2]
PeopleTec, 4901-D Corporate Drive, Huntsville, AL, USA, 35805
[1]forrest.mckee@peopletec.com      [2] david.noever@peopletec.com

## ABSTRACT

*The widespread adoption of voice-activated systems has modified routine human-machine interaction but has also introduced new vulnerabilities. This paper investigates the susceptibility of automatic speech recognition (ASR) algorithms in these systems to interference from near-ultrasonic noise. Building upon prior research that demonstrated the ability of near-ultrasonic frequencies (16 kHz - 22 kHz) to exploit the inherent properties of microelectromechanical systems (MEMS) microphones, our study explores alternative privacy enforcement means using this interference phenomenon. We expose a critical vulnerability in the most common microphones used in modern voice-activated devices, which inadvertently demodulate near-ultrasonic frequencies into the audible spectrum, disrupting the ASR process. Through a systematic analysis of the impact of near-ultrasonic noise on various ASR systems, we demonstrate that this vulnerability is consistent across different devices and under varying conditions, such as broadcast distance and specific phoneme structures. Our findings highlight the need to develop robust countermeasures to protect voice-activated systems from malicious exploitation of this vulnerability. Furthermore, we explore the potential applications of this phenomenon in enhancing privacy by disrupting unauthorized audio recording or eavesdropping. This research underscores the importance of a comprehensive approach to securing voice-activated systems, combining technological innovation, responsible development practices, and informed policy decisions to ensure the privacy and security of users in an increasingly connected world.*

## 1. INTRODUCTION

The proliferation of voice-activated systems has changed user interaction with technology, embedding these systems into everyday life. In 2024, industry forecasts suggest that the number of digital voice assistants will reach 8.4 billion units – exceeding the world's population [1]. From smartphones to smart homes, the ubiquity of voice-controlled interfaces has opened new avenues for convenience and efficiency. Recent advancements in virtual personal assistants, such as Microsoft Cortana, Apple Siri, Amazon Alexa, and Google Home, have further contributed to this revolution [2]. However, this widespread adoption has also introduced significant security and privacy concerns, particularly concerning the vulnerability of automatic speech recognition (ASR) algorithms. Researchers have demonstrated the potential for inaudible attacks on voice assistants using near-ultrasound frequencies [3-5]. These attacks can exploit the speaker to attack the microphone, potentially compromising user privacy and security. Various defensive countermeasures have been proposed to mitigate these threats, including using acoustic metamaterials to attenuate ultrasound signals [6].

In addition to inaudible attacks, researchers have explored other security and privacy concerns related to voice assistant applications. Covert channels using internal speakers have been demonstrated [7], and stealthy backdoor attacks against speech recognition systems have been proposed [8]. Surveys have highlighted the need for improved security and privacy measures in voice assistant applications [9]. Researchers have also investigated privacy-preserving hands-free voice authentication leveraging edge technology [10] and the potential for voice deepfakes to exploit digital assistants [11]. Liveness detection has been proposed as a means to enhance the security of voice assistants [12].

Furthermore, trustworthy sensor fusion techniques have been explored to defend against inaudible command attacks in advanced driver-assistance systems [13]. Comprehensive privacy and security assessments in specific voice assistants, such as Amazon Alexa, have also been conducted [14]. Ongoing research aims to secure voice processing systems from malicious audio attacks [15] and to analyze consumer IoT device vulnerability quantification frameworks [16]. Additionally, researchers have addressed the problem of accents in speech recognition systems [17], which can impact the effectiveness of voice-activated technology [18].

In this context, our research extends upon previous findings that demonstrated the potential of using near ultrasonic frequencies (16 kHz - 22 kHz) to compromise the integrity of these systems. The cornerstone of this research lies in the unique interaction between near ultrasonic noise and the microelectromechanical systems (MEMS) microphones commonly used in voice-activated devices. While designed for the human audible range, these microphones inadvertently demodulate near ultrasonic frequencies into the audible spectrum. We exploit the phenomenon to disrupt ASR algorithms. Our study explores the application of near ultrasonic noise to mitigate the effectiveness of voice-activated systems' ASR algorithms.

Our research methodology involves a systematic analysis of the impact of near ultrasonic noise on various ASR systems, considering factors such as frequency range, noise intensity, and the directional characteristics of the sound. By exploring the threshold at which these systems' performance begins to degrade, we aim to establish a comprehensive understanding of the interaction between ultrasonic frequencies and ASR algorithms. This insight is crucial for developing strategies

| Command | Distance [ft] | No Manual Trigger | Manual Trigger |
|---|---|---|---|
| Set the thermostat to fifty-five degrees | 1 | x | Ack/reply |
|  | 3 | x | Ack/reply |
|  | 6 | x | Ack/reply |
| What's the forecast for this Thursday? | 1 | Ack/reply | n/a |
|  | 3 | x | n/a |
|  | 6 | x | n/a |
| Find songs that feature flutes. | 1 | misheard | n/a |
|  | 3 | misheard | n/a |
|  | 6 | x | n/a |
| Search for the film 'The Shawshank Redemption.' | 1 | Ack | n/a |
|  | 3 | x | n/a |
|  | 6 | x | n/a |
| Show me the fastest way to sift through these files. | 1 | x | n/a |
|  | 3 | Ack | n/a |
|  | 6 | Ack | n/a |
| Play an audiobook on Audible. | 1 | x | Ack/reply |
|  | 3 | x | Ack/reply |
|  | 6 | x | Ack/reply |
| Who won the World Cup in 2018? | 1 | x | n/a |
|  | 3 | x | n/a |
|  | 6 | x | n/a |
| Remind me to buy milk tomorrow. | 1 | Ack/reply | n/a |
|  | 3 | Ack | n/a |
|  | 6 | Ack/reply | n/a |
| Turn on bedroom light. | 1 | x | n/a |
|  | 3 | Ack | n/a |
|  | 6 | x | n/a |
| Play music by Adele. | 1 | Ack/reply | n/a |
|  | 3 | x | n/a |
|  | 6 | x | n/a |

*Figure 1. The table compares the voice activate system (VAS) responses at different distances (1, 3, and 6 feet) for various spoken commands. An "Ack" or "Ack/reply" indicates the VAS acknowledged the command, while an "X" denotes no response.*

to protect voice-activated systems from malicious attacks that exploit this vulnerability.

This paper presents a detailed examination of our findings, discussing the implications for the security of voice-activated systems and offering recommendations for enhancing their resilience against such interference. Our results contribute to cybersecurity and have broader implications for designing and developing more robust and secure voice-activated technologies. We also explore the converse use case where active jamming increases voice privacy by masking ordinary room conversations and making wake-word activations less possible when users want devices disabled by their own controls.

## 2. METHODS

To investigate the vulnerability of automatic speech recognition (ASR) systems to near-ultrasonic interference, we designed a series of experiments that controlled the test environment's noise modulation, broadcasting, sound intensity, and spatial arrangement. The following methods sections detail our study's specific techniques and parameters.

**2.1 Noise Modulation and Broadcasting.** We employed white noise as the interference signal in our experimental setup. This noise was specifically modulated to near-ultrasonic frequencies using SUSBAM (Single Upper Sideband Amplitude Modulation) [3-5], aligning with our objective to examine the demodulation effect via the device's MEMS microphone. White noise was chosen due to its broad frequency spectrum, ensuring a comprehensive test of the microphone's response across the targeted ultrasonic range. Further work employing color noise (brown, pink, etc.) or random speech sometimes described as "coffee-shop noise" could make filtering interference less accessible at the device or ASR server.

**2.2 Sound Intensity Measurement.** To ensure consistency and reproducibility, we calibrated the speaker to emit the modulated white noise at a sound intensity of approximately 60 dB. This level was selected to mimic the typical sound levels in human conversation, providing a realistic context for evaluating the potential impact of ultrasonic interference on ASR systems in everyday environments. The sound intensity was measured at 3 feet from the Amazon Echo Dot (2nd Generation), which served as the primary test device for these experiments.

**2.3 Spatial Arrangement and Testing Distances.** Recognizing the potential influence of spatial variables on the effectiveness of ultrasonic interference, we positioned the audible broadcasting speaker at three distinct distances from the Echo Dot: 1 foot, 3 feet, and 6 feet. This arrangement allowed us to observe the effects of varying proximity between the noise source and the device, highlighting factors that resemble the signal-to-noise ratio and the influence of speaker distance from the microphone. By testing the ASR system's performance at these different distances, we aimed to understand the spatial dynamics involved in ultrasonic interference and their implications for the vulnerability of voice-activated systems.

## 3. RESULTS

Figure 1 summarizes the primary results of our experiment. We tested the wake and activation commands at three different distances (1 ft, 3 ft, and 6 ft) from the device, both with and without manual activation (i.e., manually triggering the device before the command). "Ack/reply" indicates that the device acknowledged the command and provided an appropriate response, "misheard" indicates an incorrect interpretation of the command, and "x" denotes no response or recognition by the device. Summarizing the critical findings shown in tabular form in Figure 1:
1. Simple commands like "Set the thermostat to fifty-five degrees" and "Play an audiobook on Audible" consistently received "Ack/reply" responses at all distances.

2. More complex or multi-part commands such as "What's the forecast for this Thursday?", "Find songs that feature flutes" and "Who won the World Cup in 2018?" had mixed results, with some distances yielding no response.
3. Commands requiring specific information, like "Search for the film 'The Shawshank Redemption,'" "Show me the fastest way to sift through these files," and "Play music by Adele," often resulted in no response at most distances.
4. Commands that emphasize the consonants "s," "f," and "th," such as those illustrated in finding three, are also known to score lower on recognition in broadly defined voice activation systems [17]. Conversely, commands lacking those consonants in finding one above have a higher Ack/reply rate.
5. The command "Turn on bedroom light" was acknowledged at 3 feet but not at other distances.
6. The command "Remind me to buy milk tomorrow" consistently received "Ack/reply" at all distances.

Overall, the results suggest that the VAS performance is affected by both the complexity of the spoken command and the distance from the user. Simple, direct commands work best across all distances, while more intricate or information-heavy requests are less reliable, especially at greater distances. Further research could explore ways to optimize ASR systems for better performance with complex commands and at various distances from the user. To visualize the effects of near-ultrasonic demodulated noise, we used voice analysis software (Audacity) to display contrasting noisy environments with spectrograms.

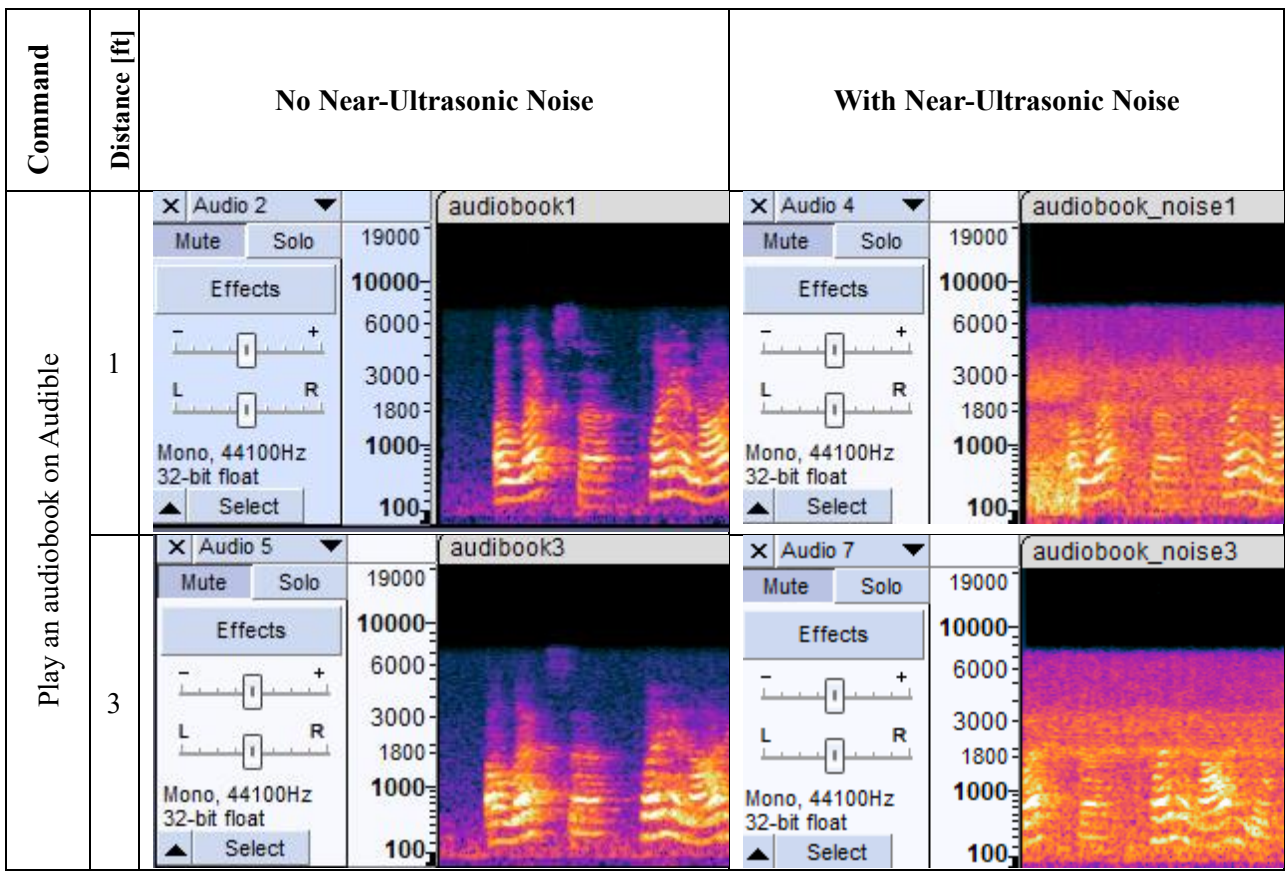

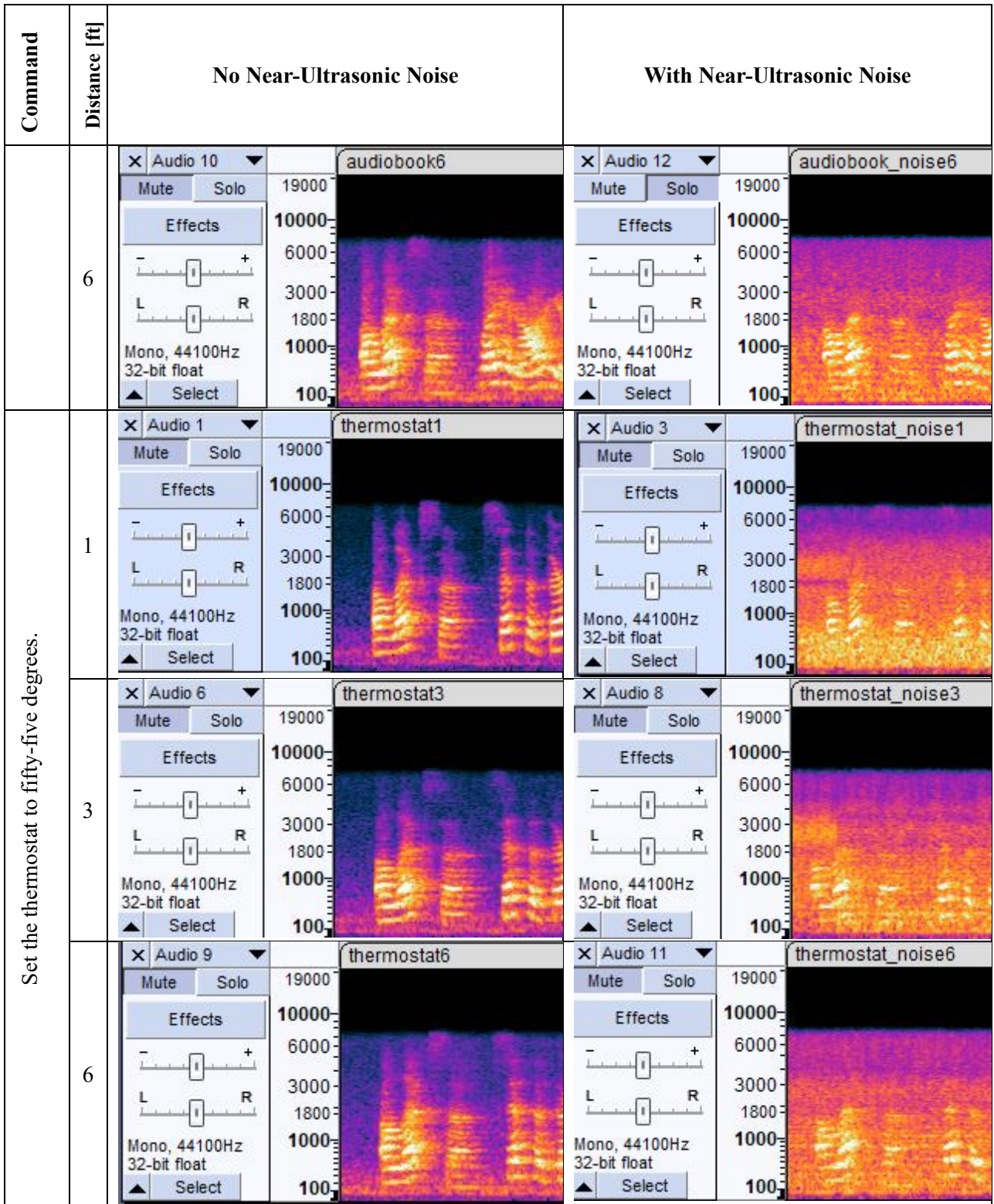

*Figure 2. Spectrograms of voice-activated commands with and without near-ultrasonic noise interference at varying distances.*

Figure 2 illustrates the impact of near-ultrasonic noise on the clarity and comprehensibility of voice-activated commands captured by the device's microphone. By presenting paired spectrograms, one without interference and one with the presence of 16-22 kHz noise, the figure demonstrates the significant distortion caused by the microphone's unintended demodulation of the inaudible portions of the signal.

Moreover, including multiple distances (1, 3, and 6 feet) in the experimental setup allows a visual comparison of how the interference pattern changes with varying proximity between the noise source and the device. This representation suggests the spatial dynamics of the vulnerability and its implications for the effectiveness of voice-activated systems in different real-world scenarios.

Notably, the spectrograms with near-ultrasonic noise interference exhibit a broad, blurred pattern that obscures the original voice command, rendering it increasingly incomprehensible to the ASR system. This visual evidence supports the hypothesis that near-ultrasonic frequencies can disrupt the performance of voice-activated devices by exploiting the MEMS microphone's demodulation of inaudible signals into the audible spectrum.

## 4. DISCUSSION

One novel aspect of these findings lies in exploiting the unintended demodulation properties of MEMS microphones, which are widely used in voice-activated devices. By leveraging this vulnerability (CVE-2023-33248 [18]), it is possible to create a 'sonic shield' that can effectively protect against unauthorized audio recording or eavesdropping. This technology has potential applications in various domains where privacy is paramount, such as sensitive business meetings, personal conversations, or situations involving minors or non-voluntary subjects.

The experimental results presented in this study demonstrate the significant impact of near-ultrasonic noise on the performance of automatic speech recognition (ASR) systems. Figure 1 summarized the task completion rates for various voice-activated commands drastically decrease in the presence of interfering ultrasonic noise compared to the control scenario without interference. This finding highlights the vulnerability of commercial ASR microphones to inaudible noise in the 16-22 kHz range, which can effectively disrupt the device's ability to process and interpret voice commands accurately.

Figure 2 explored the underlying mechanism of this interference by visualizing the spectral characteristics of the voice commands with and without the presence of ultrasonic noise. The spectrograms illustrate the demodulation of the inaudible noise by the microphone, resulting in a broad, blurred pattern that obscures the original voice command. This visual evidence supports the hypothesis that near-ultrasonic frequencies can significantly degrade the audio signal quality prior to speech-to-text conversion, rendering the voice command incomprehensible to the ASR system.

### 4.1 Covert Communication

One potential application of this technology is in the realm of covert communication. By deploying near-ultrasonic noise, it is feasible for individuals to communicate verbally in the presence of potential eavesdropping devices without being overheard or recorded. This method capitalizes on the inability of most commercial microphones, specifically those used in popular voice-activated devices, to effectively process speech signals embedded in or masked by near-ultrasonic noise. Such a technique could be invaluable in sensitive environments where privacy is paramount, such as in diplomatic meetings, corporate boardrooms, or journalistic investigations.

### 4.2 Privacy and Security Implications

The ability to generate a near-ultrasonic noise field to mask conversations opens new possibilities for privacy protection in an era of pervasive audio surveillance. Few encounters arise in which one or both parties lack always-listening devices, whether smartphones, wearables such as watches or eyeglasses or internet-of-things home automation devices. By deploying this jamming technology as a wearable device or a room-based system, individuals can ensure that their conversations remain private and unrecorded by nearby voice-activated devices or eavesdropping equipment. This technique could enable journalists, activists, or whistleblowers to communicate securely and protect their sources and information.

Moreover, integrating near-ultrasonic noise generation into voice-activated systems themselves could provide an additional layer of privacy protection. These devices can effectively prevent unintentional or unauthorized recording of ambient conversations by continuously broadcasting inaudible noise and filtering it out before the wake word detection stage. This approach could help address the growing concerns surrounding the privacy implications of always-listening voice assistants in homes and workplaces.

### 4.3 Risks and Misuse Potential

While the potential applications of this technology for privacy protection are promising, it is crucial to acknowledge the risks and ethical considerations associated with its use. The same technique that can be used to safeguard privacy could also be employed maliciously to disrupt legitimate voice-activated systems, such as those used for emergency assistance or accessibility purposes. When individuals rely on these systems for critical communication or support, the interference caused by near-ultrasonic noise could have consequences. Furthermore, the development of this technology raises questions about the potential arms race between privacy protection measures and surveillance capabilities. As researchers and engineers find new ways to shield conversations from unauthorized recording, those seeking to overcome these barriers may respond with more advanced eavesdropping techniques. These methods underscores the need for ongoing research and dialogue to ensure that privacy protection technologies remain ahead of the curve while minimizing the potential for misuse.

## 5. CONCLUSIONS AND FUTURE WORK

In conclusion, our research has expanded upon a critical vulnerability in voice-activated systems [18]: the susceptibility of their automatic speech recognition (ASR) algorithms to near-ultrasonic noise interference. The study's analysis underscores that the MEMS microphones, a common component in these systems, demodulate near-ultrasonic frequencies into the audible range, disrupting the ASR process.

The susceptibility of voice-activated systems to ultrasonic interference presents a novel avenue for malicious exploitation, posing a threat to individual users and organizations. However, this research also highlights the potential for harnessing this vulnerability to protect privacy by actively disrupting unauthorized audio recording or eavesdropping.

Developing devices and software solutions that leverage near-ultrasonic or ultrasonic interference could provide individuals with additional control over their audio privacy. For instance, future work could explore creating mobile applications that continuously broadcast inaudible noise to prevent unintentional recording by nearby voice-activated devices. Similarly, integrating acoustic damping materials or active interference capabilities into phone cases or covers for smart home devices could offer users a physical means of protecting their conversations from unwanted surveillance.

Moreover, our findings underscore the need for a multifaceted approach to securing voice-activated systems against ultrasonic interference. While developing privacy-enhancing technologies is essential, it is equally important for the industry to invest in research to create more resilient ASR algorithms and microphone designs less prone to ultrasonic interference. This goal may involve exploring advanced signal processing techniques, machine learning algorithms, or novel hardware architectures that can effectively filter out or mitigate the impact of near-ultrasonic noise.

### ACKNOWLEDGMENTS
The authors thank the PeopleTec Technical Fellows program for encouragement and project assistance.